\documentclass[fleqn,usenatbib]{mnras}

\usepackage[T1]{fontenc}
\usepackage{ae,aecompl}

\newcommand{\kms}{$\mbox{km~s}^{-1}$}

\bibpunct{(}{)}{;}{a}{}{,}

\def\apj{{ApJ}}                 
\def\apjl{{ApJ}}                
\def\apjs{{ApJS}}               
\def\ao{{Appl.~Opt.}}           
             
\def\aap{{A\&A}}

\def\mnras{{MNRAS}}

\def\pasa{{PASA}}

\usepackage{comment}
\usepackage{color,epsfig}

\usepackage{graphicx}	
\usepackage{amsmath}	
\usepackage{amssymb}	

\title[Ammonia masers]{Ammonia masers toward G\,358.931$-$0.030}

\author[T. P. McCarthy et al.]{T. P. McCarthy,$^{1}$\thanks{E-mail: tiegem@utas.edu.au}
S. L. Breen,$^{2}$
J. F. Kaczmarek,$^{3,4}$
X. Chen,$^{5,6}$
S. Parfenov,$^{7}$
\newauthor A. M. Sobolev,$^{7}$
S. P. Ellingsen,$^{1}$
R. A. Burns,$^{8}$
G. C. MacLeod,$^{9,10}$
K. Sugiyama,$^{11}$
\newauthor
A. L. Brierley$^{12}$
and S. P. van den Heever$^{10}$
\\
$^{1}$School of Natural Sciences, University of Tasmania, Private Bag 37, Hobart, Tasmania 7001, Australia\\
$^{2}$SKA Observatory, Jodrell Bank, SK11 9FT, UK\\
$^{3}$CSIRO Astronomy and Space Science, Australia Telescope National Facility, Box 76, Epping, NSW 1710, Australia\\
$^{4}$Dominion Radio Astrophysical Observatory, National Research Council of Canada, Box 248, Penticton, BC, V2A 6J9, Canada\\
$^{5}$Center for Astrophysics, GuangZhou University, Guangzhou 510006, People's Republic of China\\
$^{6}$Shanghai Astronomical Observatory, Chinese Academy of Sciences, Shanghai 200030, People's Republic of China\\
$^{7}$Ural Federal University, 19 Mira street, 620002 Ekaterinburg, Russia\\
$^{8}$Department of Science, National Astronomical Observatory of Japan, 2-21-1 Osawa, Mitaka, Tokyo 181-8588, Japan\\
$^{9}$The Open University of Tanzania, P.O. Box 23409, Dar-Es-Salaam, Tanzania\\
$^{10}$SARAO, Hartebeesthoek Radio Astronomy Observatory, PO Box 443, Krugersdorp, 1741, South Africa\\
$^{11}$National Astronomical Research Institute of Thailand (Public Organization), 260 Moo 4, T. Donkaew, A. Maerim, Chiangmai, 50180 Thailand\\
$^{12}$Crompton House}

\date{Accepted XXX. Received YYY; in original form ZZZ}

\pubyear{2022}

\begin{document}
\label{firstpage}
\pagerange{\pageref{firstpage}--\pageref{lastpage}}
\maketitle

\begin{abstract}
We report the detection of ammonia masers in the non-metastable (6, 3), (7, 5) and (6, 5) transitions, the latter is the first unambiguous maser detection of that transition ever made. Our observations include the first VLBI detection of ammonia maser emission, which allowed effective constrain of the (6, 5) maser brightness temperature. The masers were detected towards G\,358.931$-$0.030, a site of 6.7-GHz class~II methanol maser emission that was recently reported to be undergoing a period of flaring activity. These amnmonia masers appear to be flaring contemporaneously with the class~II methanol masers during the accretion burst event of G\,358.931$-$0.030. This newly detected site of ammonia maser emission is only the twelfth such site discovered in the Milky Way. We also report the results of an investigation into the maser pumping conditions, for all three detected masing transitions, through radiative transfer calculations constrained by our observational data. These calculations support the hypothesis that the ammonia (6, 5) maser transition is excited through high colour temperature infrared emission, with the (6, 5) and (7, 5) transition line-ratio implying dust temperatures $> 400$~K. Additionally, we detect significant linearly polarised emission from the ammonia (6, 3) maser line. Alongside our observational and radiative transfer calculation results, we also report newly derived rest frequencies for the ammonia (6, 3) and (6, 5) transitions.

\end{abstract}




\section{Introduction}

Ammonia masers have now been detected in at least 32 metastable ($J = K$) and non-metastable ($J > K$)  transitions, plus a further two isotopic ammonia transitions. The first detections of probable weak masers in the (3, 3) line were reported towards W33 and DR 21(OH) \citep{Wilson-1982,Guilloteau-1983}. The unambiguous detection of non-metastable ammonia transitions were made soon after in the (9, 6) transition, initially serendipitously during observations targeting NH$_2$CCH. Subsequent observations detected (9, 6) masers towards W51, W49, DR 21(OH), and NGC 7538 as well as a (6, 3) maser towards W51 \citep{Madden-1986}. Detailed ammonia studies of W51 followed, quickly revealing additional ammonia masers in the (7, 5), (11, 9), (5, 4) and (9, 8) transitions \citep[][(10, 8) was also considered a probable maser detection but was not confirmed in later observations by Henkel et al. (2013)]{Wilson-1988,Mauersberger-1987}, (3, 3) \citep{Zhang-1995} and then in the (6, 2), (5, 3), (7, 4), (8, 5), (6, 6), (7, 6), (8, 6), (7, 7), (9, 7) (10, 7), (9, 9), (10, 9) and (12, 12) transitions \citet{Henkel-2013}. NGC\,7538 also proved to be a prominent site of ammonia maser emission with further studies revealing (10, 6), (10, 8) and (9, 8) masers \citep{Hoffman-2011,Hoffman-2011b,Hoffman-2012,Hoffman-2014}, as well as $^{15}$NH$_3$ masers in the (3, 3) and (4, 3) transitions \citep{Schilke-1991,Mauersberger-1986,Johnston-1989}. Ammonia (3, 3) masers have also subsequently been detected towards DR 21(OH) \citep{Mangum-1994}.

Since the early detections only a handful of further ammonia maser sites have been discovered: NGC\,6334 in the (11, 9), (8, 6), (7, 6), (9, 9), (10, 9) \citep{Walsh-2007}, (6, 6) \citep{Beuther-2007}, and (3, 3) \citep{Kraemer-1995} transitions; Sgr\,B2 in the (2, 2) \citep{Mills-2018}, (9, 3), (9, 4), (8, 3), (9, 5), (8, 4), (7, 3), (6, 1) and (6, 4) transitions \citep{Mei-2020}; IRAS20126+4104 in (3, 3) \citep{Zhang-1999}; G\,5.89$-$0.39 in (3, 3) \citep{Hunter-2008}; G\,9.62+0.19 in (5, 5) \citep{Hofner-1994}; G\,23.33$-$0.30 in (3, 3) \citep{Walsh-2011}; and G\,19.61$-$0.23 in (11, 9) and (8, 6) \citep{Walsh-2011}. 

The remarkably small number of ammonia maser sites is despite the extensive, 100 square degree survey of the southern Galactic plane for ammonia lines in the (8, 6), (9, 7), (7, 5), (11, 9), (4, 1), (1, 1), (2, 2), (3, 3), (6, 6) and (9, 9) transitions as part of HOPS \citep[H$_2$O Southern Galactic Plane Survey][]{Walsh-2011}. HOPS has a modest detection limit (with typical 3$\sigma$ values of between 3 to 6~Jy) but provides evidence that strong ammonia masers in these transitions are rare.

High spatial resolution observations have revealed the close spatial coincidence between the detected masers and regions of shocks, often found at the interface of molecular outflows with the surrounding environment \citep[e.g.][]{Kraemer-1995,Zhang-1995}. The pumping of both metastable and non-metastable ammonia lines is still poorly understood. Currently, the best model for non-metastable maser population inversion is based on vibrational excitation via $10\mu$m infrared radiation\citep{Brown-1991}. This pumping mechanism is further supported by observation of maser pairs in the $K$ ladders for ammonia transitions (i.e. $J$ and $J+1$ for the same $K$ value) by \citep{Henkel-2013} toward W51-IRS2. Variability studies of the ammonia maser lines in W51 revealed moderate levels of temporal intensity changes and further that the (9, 6) maser line remained stronger than the (7, 5) and (6, 3) lines \citep{Wilson-1988}. In other monitoring observations, no obvious temporal variability was identified \citep{Hoffman-2011b}.

Here we present a series of ammonia observations, including a number of metastable and non-metastable transitions, towards the location of a 6.7-GHz class II methanol maser, G\,358.931$-$0.030, discovered in the Methanol Multibeam survey \citep[MMB;][]{Caswell-2010}. The observations were conducted during a period of 6.7-GHz methanol maser flaring activity which was identified by a global cooperative of maser monitoring programs called `M2O', and first reported by \citet{Coconuts-2019}. This flaring event has been attributed to an accretion burst event from the high-mass protostar associated with these masers \citep{Burns-2020, Stecklum-2021}. The flaring 6.7-GHz methanol maser has provided an interesting site for multi-wavelength followup observations, including searches for new maser transitions, subsequent VLBI observations and temporal monitoring. To date the discovery of 22 new class II methanol lines have been reported, including the detection of fifteen torsionally excited lines \citep{Breen-2019,Brogan-2019,MacLeod-2019}, and the first detection of isotopic methanol masers in three transitions \citep{Chen-2019}. This large variety of rare maser species detected during the flaring event was the motivation behind this search for ammonia lines toward G\,358.931$-$0.030.

\section{Observations and Data reduction} \label{sec:obs}

\begin{table*}

 \caption{Details of the primary interferometric ammonia observations, including telescope, array configuration (when the full array was unavailable, the antennas used are given in parenthesis), the minimum and maximum baseline lengths, the observation bandwidths used to cover target spectral lines (multiple values indicate different bandwidths were used for different zoom bands), frequency resolution, phase calibrator used, along with the total time of the observations and the total integration time.} 
  \begin{tabular}{llccccccl} \hline
 \multicolumn{1}{l}{\bf Telescope} &\multicolumn{1}{c}{\bf Epoch}  	& \multicolumn{1}{c}{\bf Array} 	& \multicolumn{1}{c}{\bf Min, Max} &		  \multicolumn{1}{c}{\bf BW} & \multicolumn{1}{c}{\bf res.} &	\multicolumn{1}{c}{\bf Phase}&\multicolumn{1}{c}{\bf Total time} & \multicolumn{1}{c}{\bf Int. time}\\
 &  \multicolumn{1}{l}{\bf DOY}
 & & \multicolumn{1}{c}{\bf baseline (m)}&\multicolumn{1}{c}{\bf (MHz)}  & \multicolumn{1}{c}{\bf (kHz)}		& \multicolumn{1}{c}{\bf calibrator} 	
  	 & \multicolumn{1}{c}{\bf (hours)} & \multicolumn{1}{c}{\bf (hours)}  \\ \hline
ATCA & 064 & H214(3,4,5) & 82, 240 &64 & 32 & B1741$-$312 & 4 & 0.8 \\
ATCA & 084 & 6A & 336, 5938 &3.5 & 0.5 & B1714$-$336 &  4.75 & 3\\
VLA & 094 & B & 210, 11100 &4 & 8 & J1744$-$3116 & 1.5 & 0.5\\
ATCA & 096 & H75 & 30, 89 &2, 2.5, 3 & 0.5 & B1714$-$336 & 6 & 3.6 \\
ATCA & 101 & 750C & 45, 5020&2 & 0.5 & B1714$-$336 & 5.5 & 2.4 \\
\hline
\end{tabular}\label{tab:obs}
\end{table*}

\begin{table*}
 \caption{Target ammonia lines, followed by the adopted rest frequency (with errors in the last digit in parenthesis) and observation epoch preceded by a letter indicating the array (A=ATCA, V=VLA), smoothed to the resolution shown. Detected transitions are marked with a leading `$^*$'. Rest frequencies marked with a trailing `$\dag$', rather than numerical reference, are those from our calculations which deviate from the previously reported values in the literature (see Section \ref{sec:restfreq}).} 
 \begin{tabular}{rlccccccl} \hline
 \multicolumn{1}{c}{\bf Spectral} &\multicolumn{1}{c}{\bf Rest freq.}  	& \multicolumn{1}{c}{\bf 2019 Epoch} 	&		  \multicolumn{1}{c}{\bf V$_{coverage}$} & \multicolumn{1}{c}{\bf V$_{res.}$} &	\multicolumn{1}{c}{\bf Beam}&\multicolumn{1}{c}{\bf RMS}\\
  \multicolumn{1}{c}{\bf line}
 &\multicolumn{1}{c}{\bf (MHz)}  &		\multicolumn{1}{c}{\bf (telescope DOY)} 	
  	 & \multicolumn{1}{c}{\bf (\kms)} & \multicolumn{1}{c}{\bf (\kms)} &	\multicolumn{1}{c}{\bf ($''$ $\times$ $''$)}& \multicolumn{1}{c}{\bf (mJy)}\\ \hline

NH$_3$ (9, 6) & 18499.390(5)$^{[1]}$  & A101 & 32 & 0.02 & 14.6$\times$2.4 & 24.8\\
NH$_3$ (8, 5) & 18808.507(5)$^{[2]}$  & A096  & 39  & 0.02  &  27.1$\times$21.2 & 31.5 \\
 &   & A101 &  31 & 0.02 &  14.3$\times$23.3 & 33.6 \\

$^*$NH$_3$ (6, 3) & 19757.579(10)$^{\dag}$  & V094  & 60  & 0.12 & 1.62$\times$0.26 & 80 \\
 &  & A096  & 30 & 0.02 & 26.0$\times$20.5 &  38.9\\
NH$_3$ (8, 6) & 20719.221(10)$^{[2]}$  & A064  & 925 & 0.45 & 10.7$\times$4.7 & 33.4 \\
 &  & A096 & 36 & 0.02 & 25.3$\times$19.5 & 50.0\\
{$^*$NH$_3$ (7, 5)} & 20804.830(5)$^{[2]}$ & A064  & 922 & 0.45  & 10.6$\times$4.7 & 24.5 \\
 &   & A084  &  50  &  0.02  & 1.77$\times$0.39  & 70.0 \\
 &   & V094 & 57 & 0.11 & 1.20$\times$0.32 & 80 \\
NH$_3$(11, 9)& 21070.739(5)$^{[3]}$  & A064 & 910 & 0.44 & 10.5$\times$4.6 & 27.3\\  
NH$_3$ (4, 1) & 21134.311(5)$^{[2]}$  & A064 & 907 & 0.44 & 10.5$\times$4.6 & 27.6\\ 
$^*$NH$_3$ (6, 5) & 22732.425(4)$^{\dag}$  & V094 & 52 & 0.10 & 1.06$\times$0.34 & 80\\
NH$_3$ (2, 1) & 23098.8190(1)$^{[4]}$  & A064 & 830 &  0.41 &  9.6$\times$4.2 & 35.9 \\
NH$_3$ (9, 8) & 23657.471(5)$^{[4]}$ & A096  & 31 & 0.02 & 22.1$\times$17.1 & 49.7 \\
NH$_3$ (4, 4) & 24139.4169(1)$^{[5]}$ & A096 & 31 & 0.1 & 22.7$\times$16.8 & 3.1\\
NH$_3$ (6, 6) & 25056.025(5)$^{[5]}$ & A064 & 765 & 0.37 & 8.8$\times$3.9 & 29.5 \\
\hline
\end{tabular}\label{tab:lines}
	\begin{flushleft}
	Note: $^{[1]}$\citet{Madden-1986}, $^{[2]}$\cite{Hermsen+1988}, $^{[3]}$\citet{Mauersberger-1987}, $^{[4]}$\citet{Moran-1973} and $^{[5]}$\citet{Barrett-1977}. \\
	\end{flushleft}	
\end{table*}

We report on a series of observations, including both metastable and non-metastable ammonia transitions, targeting the 6.7-GHz methanol maser G\,358.931$-$0.030 that was recently reported to be undergoing a period of flaring \citep{Coconuts-2019}. In addition to primary observations conduced with the Australia Telescope Compact Array (ATCA) and NSF's Karl G. Jansky Very Large Array (VLA), further observations have also been conducted with the 65-m Tianma (TMRT), and the 26-m Hartebeesthoek radio telescopes. The TMRT notably made the first detection of the (6, 3) and (6, 5) masers on 2019 March 17 following the initial (7, 5) ammonia maser detection made with the ATCA on 2019 March 05. Together with several epochs of Hartebeesthoek observations, the TMRT observations are included here to provide an assessment of the temporal variability of the detected ammonia maser emission. Additionally, results from a Korean VLBI Network (KVN) fringe test are included, as the detected (6, 5) emission is useful for constraining the brightness temperature of the flaring component.

Observations of both metastable and non-metastable ammonia transitions were made across four epochs with the Australia Telescope Compact Array (ATCA), and one with the NSF's Karl G. Jansky Very Large Array (VLA). These observations included 15 ammonia transitions, together with one isotopic ammonia transition, which have rest frequencies between 18 and 26~GHz (specifically, ammonia (9, 6), (8, 5), (6, 3), (8, 6), (7, 5), (11, 9), (4, 1), (6, 5), (2, 1), (9, 8), (1, 1), (2, 2), (3, 3), (4, 4), (6, 6) and $^{15}$NH$_3$ (5, 5)). Each of the transitions were observed at up to three epochs. All observations targeted the reported position of the 6.7-GHz methanol maser \citep[J2000 position: 17$^h$43$^m$10.02$^s$ $-$29$^\circ$51$'$45.8$''$;][]{Caswell-2010}, which has since been slightly refined \citep[17$^h$43$^m$10.10$^s$ $-$29$^\circ$51$'$45.5$''$;][]{Breen-2019}. Table~\ref{tab:obs} summarises the basic details of the observations, including telescope and array configuration, observation bandwidths and channel spacing, phase calibrator used, duration of the observations and total integration time. Details of the observed lines, adopted rest frequencies, observed velocity coverage, velocity resolution, synthesised beam size and RMS noise are given in Table~\ref{tab:lines}

\subsection{Rest frequency calculation} \label{sec:restfreq}

It is important to note that, for the detected ammonia maser transitions in this study, we adopted rest frequencies calculated as a weighted average over hyperfine components. The weights were the hyperfine line catalogue intensities. The hyperfine component frequencies and the catalogue intensities were taken from the VASTEL database\footnote{\url{http://cassis.irap.omp.eu/?page=catalogs-vastel}}. The rest frequencies we calculate for these lines are, 19757.579 MHz (6, 3), 20804.830 MHz (7, 5) and 22732.425 MHz (6, 5). The calculated (6, 3) and (6, 5) rest frequencies are 0.041 MHz higher and 0.004 MHz lower respectively than those previously reported in the literature \citep{Hermsen+1988, Nystrom-1978}, while the (7, 5) rest frequency is identical \citep{Hermsen+1988}.


\subsection{ATCA observations}\label{ATCA}

Four epochs of ATCA observations were conducted between 2019 March 5 and April 11 (DOY 064, 084, 096 and 101) during allocations of maintenance and Director's time. A summary of the observation specifications are given in Table~\ref{tab:obs}. Four different array configurations were used, and during March 5 only 3 of the antennas were available for science, as noted. The Compact Array Broadband Backend \citep[CABB;][]{Wilson-2011} was configured in CFB~64M-32k on March 5 and CFB 1M-0.5k for the other epochs. The former provided 64~MHz spectral zoom bands, each with 2048 channels, whereas the latter provided a number of 1~MHz zoom bands, each with 2048 channels (which have been combined in order to provide adequate velocity coverage; indicated in column 4 of Table~\ref{tab:lines}). Observations of the 6.7-GHz methanol maser target were interleaved with observations of a nearby phase calibrator (given in column 7 of Table~\ref{tab:obs}) every $\sim$8 - 10 mins. Pointing observations were made on the phase calibrator (B1741$-$312 or B1714$-$336) once per hour. At each of the four epochs, observations of PKS\,B1934$-$638 and PKS\,B1253$-$055 were made for primary flux density and bandpass calibration, respectively.

All data were reduced following the procedure outlined in \citet{Breen-2019} and details of the velocity coverage, velocity resolution (sometimes after smoothing), synthesized beam and RMS noise characteristics are given in Table~\ref{tab:lines}. Well calibrated, high SNR, ATCA observations during good weather conditions have a nominal astrometric accuracy of approximately 0\farcs5 \citep{Caswell-1997}.

\subsection{VLA observations}\label{VLA}

VLA observations of the three NH$_3$ lines were made on 2019 April 4 (DOY 094) during a 1.5 hour allocation of Director's Discretionary Time in B-array. Details of the observations are given in Table~\ref{tab:obs}. Each of the observed lines (indicated in Table~\ref{tab:lines}) was allocated a 4 MHz zoom band, each with dual polarisation and 512 spectral channels. 3C286 was used to calibrate both the bandpass and primary flux density and observations of phase calibrator J1744$-$3116 were made through out the observation. Astrometric uncertainty for our VLA B array observations is approximately 30 milliarcseconds.

Data were reduced by the VLA Calibration Pipeline using the Common Astronomy Software Applications (CASA) package, applying standard techniques for the reduction of VLA spectral line data. The {\sc miriad} task IMFIT was then used to derive the maser spot distribution by fitting a 2D Gaussian to the emission present in each channel map. This method allowed us to estimate both the position (with typical fitting errors of 0.01 arcsec) and the flux densities of the maser emission.


\subsection{Shanghai 65m Tianma Radio Telescope (TMRT) observations}\label{TMRT}

Detetections of the (6, 3), (7, 5) and (6, 5) ammonia transitions were made during TMRT observations conducted on 2019 March 17 (DOY 076). The observations used a cryogenically cooled K-band receiver and the digital backend system (DIBAS), allowing each line a 23.4 MHz spectral band, each with 4096 channels (corresponding to a velocity channel spacing of at least 0.09~\kms). Observations of the target were followed by reference observations with a 30 second integration time on each. The total on-source time was 22 minutes. The noise diode was used to carry out flux density calibration, and is expected to be accurate to $\lesssim\,10\%$

At the frequency of these observations, the TMRT has a beamsize of $\sim$45 arcseconds and an aperture efficiency of $\sim$50 per cent, corresponding to a sensitivity of 1.6 Jy K$^{-1}$. During the observations, the system temperature was between 100--150 K, resulting in a typical RMS noise of 0.15~Jy per spectral channel.






\subsection{Hartebeesthoek observations}\label{Hart}

Monitoring observations of the (7, 5) and (6, 5) transitions were also conducted with the Hartebeesthoek radio telescope beginning on 2019 March 27 (DOY 086), and ending on 2019 April 3 (DOY 093) and 2019 August 3 (DOY 215) for the (7, 5) and (6, 5) transitions respectively. The 1.3 cm receiver utilised for these observations is a cryogenically cooled, dual-polarisation receiver. The point source sensitivity values for RCP and LCP are 10.79 and 10.36 Jy/K respectively. The velocity extent of the observations, for each of the observed transitions, is -46 to +12 km/s with a resolution of 0.1 km/s and a beamsize of ~2.2 arcmin. Flux density calibration of the Hartebeesthoek observations are expected to be accurate to $\lesssim\,10\%$.

\subsection{Korean VLBI Network observations}\label{KVN}

Two hours of observing time were approved with the KVN in response to a request for Directors Discretionary Time for follow-up imaging of the maser flare event. Observations were conducted on the 25th of March 2019 (DOY 084) with three stations. Data were recorded at an 8 Gbps recording rate, providing 512 MHz bandwidth of left-hand circular polarised signal at each of the four simultaneously operating K, Q, W and D frequency bands of the KVN, with 2-bit, Nyquist sampling. Data were correlated at the Korea Japan Correlation Center (KJCC; \citealt{Lee15a}) with a frequency channel separation of 15.625 kHz, corresponding to 0.21 km s$^{-1}$ velocity spacing. The resulting beam size for these KVN observations is $17.3 \times 1.9$~milli-arcseconds.

Two minute scans of G358 were interspersed with 2 minute scans of quasar J1744-3116 which was used for phase calibration and the observing session began and ended with scans on delay calibrators NRAO530 and BLLAC. Data were reduced using the Astronomical Image Processing Software (\emph{AIPS}, \citealt{Greisen03}) in which flux calibration was carried out using system noise measurements and gain characteristics of each of the three stations, providing a flux determination that is expected to be accurate to $\sim 10 \%$.

\section{Results} \label{sec:results}

Observations of 15 ammonia transitions and one isotopic ammonia transition have resulted in the identification of a further rare site of ammonia maser emission detected in our Galaxy. In total, three ammonia maser lines were detected towards G\,358.931$-$0.030; the (6, 3), (7, 5) and (6, 5) non-metastable transitions, the latter being the first unambiguous astronomical maser ever detected in that transition. The characteristics of each of the detections made with the ATCA and VLA are summarised in Table~\ref{tab:detections}, including the fitted position of the maser peak emission as well as flux density and velocity information. Observations of the (6, 3) and (7, 5) transitions made with the ATCA include full polarisation information and we find significant linearly polarised emission in the stronger (6, 3) transition but no significant circularly polarised emission (i.e. greater than 0.5\% of the Stokes I flux density). Higher sensitivity observations of the (7, 5) transition would be needed to confidently determine if there are low levels ($>$2$\%$ of the Stokes I flux density) of linear or circular polarised emission. 

\begin{figure}\hspace{-2.7cm}
\epsfig{figure=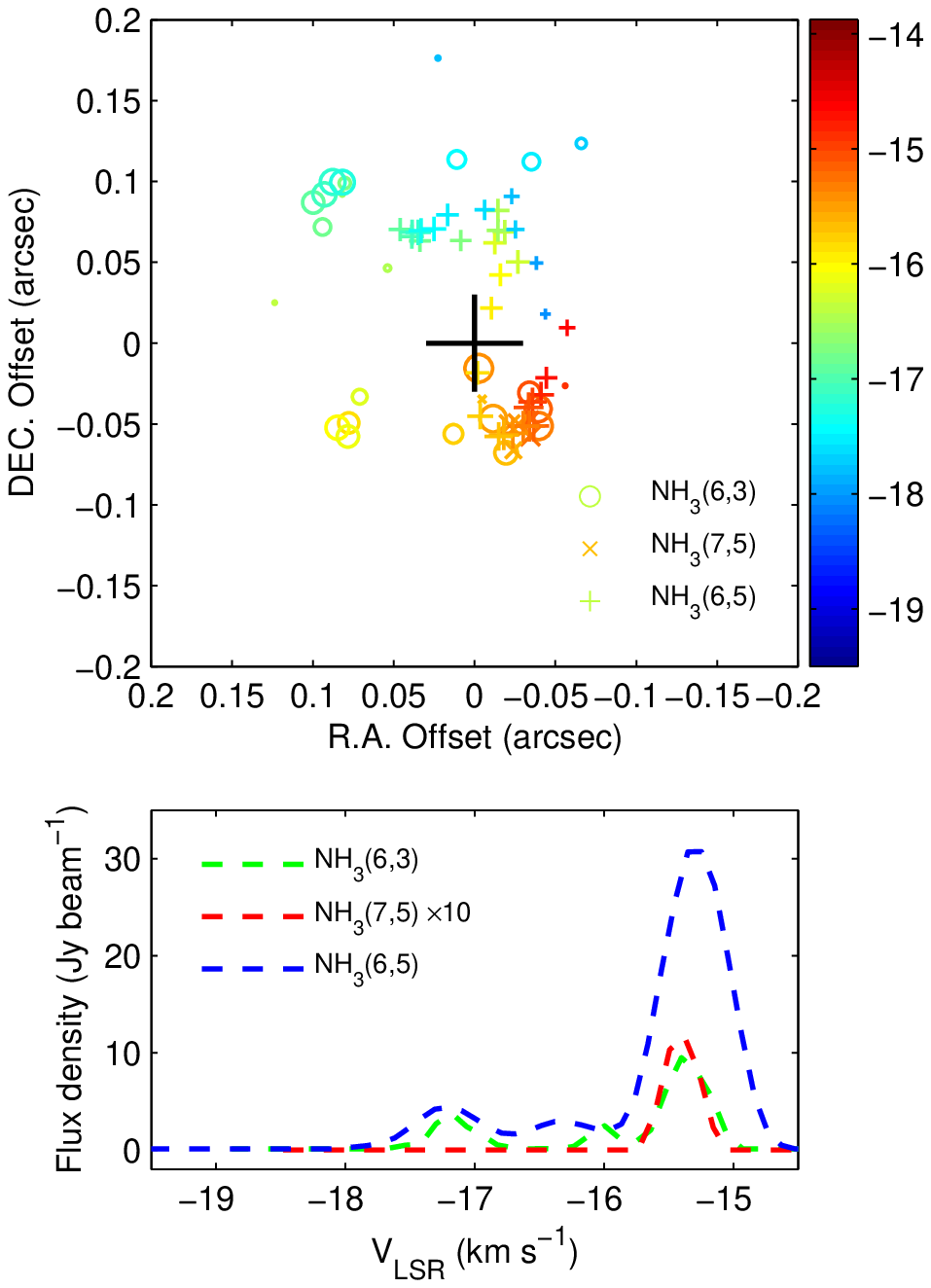,scale=0.78}\vspace{-1cm}
\caption{VLA ammonia (6, 3), (7, 5) and (6, 5) maser spot distribution relative to ALMA continuum source MM1 (J2000 position: $\alpha$=17:43:10.1015, $\delta$=-29:51:45.6936) marked by a black cross. The extent of the black cross shows the absolute positional uncertainty of 0\farcs03, while the extent of the coloured symbols indicates the positional uncertainty in the individual maser spot.
}
\label{fig:vla}
\end{figure}

Absolute positions of the detected ammonia masers (Table~\ref{tab:detections}) confirm that they are associated with the target star-formation region, which was displaying contemporaneous flaring in the 6.7-GHz class~II methanol maser transition. A detailed maser spot map, showing the distribution of the respective maser lines is given in Fig.~\ref{fig:vla}. This figure shows that the peak emission (at the least negative velocities) in each of the transitions is located to the east of the continuum source, while the bulk of the additional (6, 3) and (6, 5) emission is more widely distributed, mostly to the north.  

\begin{figure}
\epsfig{figure=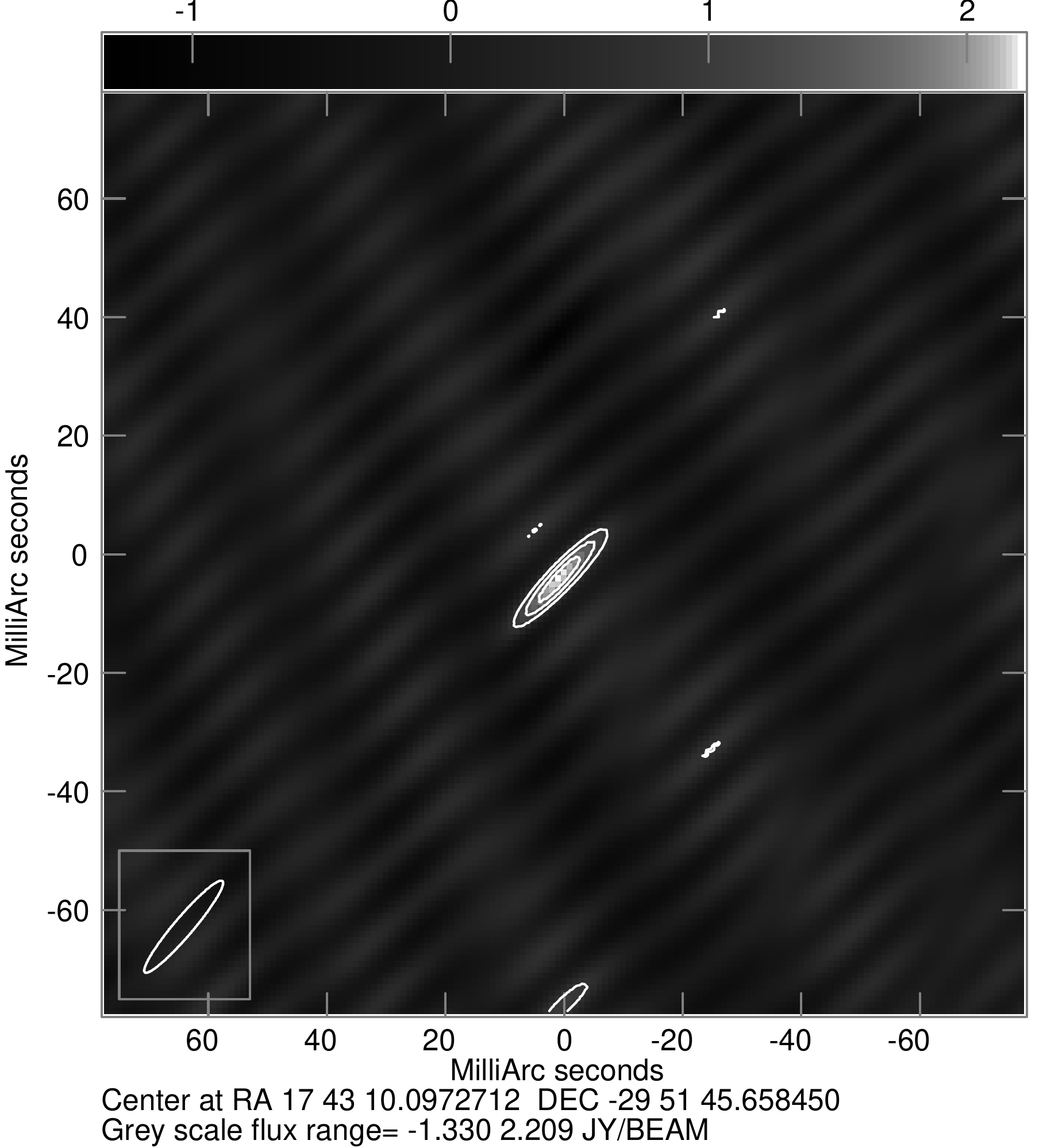,scale=0.44}
\caption{Field image of the NH$_3$ (6, 5) emission as detected during the KVN fringe test. Flux densities are shown in grey-scale and contours mark the 3, 5, 7 and 9~$\sigma$ levels ($\sigma = 0.26$~Jy\,beam$^{-1}$). The synthesized beam shape for the experiment is shown in the plot bottom-left.
}
\label{fig:kvn}
\end{figure}

Ammonia (6,5) was detected in the autocorrelation and cross-correlation KVN data, and while the total integration time was not sufficient to conduct accurate astrometry, the KVN was able to produce an image of the emission at milliarcsecond resolution (Figure \ref{fig:kvn}) in order to help constrain brightness temperatures. The image shows compact, unresolved maser emission with an integral flux of $2.12 \pm 0.41$ Jy emitting from a region with a deconvolved angular full-width half-maximum size of $17 \times 3$ milliarcseconds, determined by a 2D Gaussian fit to the VLBI image. The $2.38 \pm 0.25$ Jy/beam flux density of the maser corresponded to a >9$\sigma$ detection against the 0.253 Jy/beam RMS noise of the processed VLBI image.

Spectra for each of the maser detections are given in Fig.~\ref{fig:spect} and include a number of observation epochs. In the case of the (6, 3) maser transition, the linearly polarised emission (scaled by a factor of 10) detected with the ATCA on DOY096 is shown in addition to the Stokes I emission. Despite the relatively large velocity uncertainty (related to the uncertainty in the line rest frequency; listed in Table~\ref{tab:detections} and shown in Fig.~\ref{fig:spect}), all of the the ammonia maser features are contained within the velocity range of the target 6.7-GHz methanol maser \citep[$-$19.6 to $-$12.7~\kms;][]{Breen-2019}.
 
Significant changes in the detected spectral profiles are seen (Fig~\ref{fig:spect}) across the observation epochs for all three ammonia transitions. The first ammonia maser detection was made on DOY064 in the (7, 5) transition, just 50 days after the 6.7-GHz methanol maser was reported to be undergoing flaring activity \citep{Coconuts-2019}.  The peak flux density of the 6.7-GHz methanol maser on DOY065 was 981 Jy \citep{Breen-2019}. Our observations indicate that the period of ammonia maser emission is likely to have been a short, transient phenomenon, closely linked to the 6.7-GHz methanol maser flaring activity, with both the (7, 5) and (6, 5) transitions appearing to wane significantly, first in the (7, 5) transition. Though outside the range of Figure~\ref{fig:spect}, further monitoring from Hartebeesthoek finds the (6, 5) transition is detectable up until DOY133, with no detection of the emission in the following epoch on DOY139.

In addition to these maser lines we have detected thermal emission in the (1, 1), (2, 2) and (3, 3) metastable transitions of ammonia. 

\begin{table*}
 \caption{Properties of the ammonia masers detected towards G\,358.931$-$0.030 derived from ATCA and VLA observations, including the measured peak positions, the minimum, maximum and peak velocity, uncertainty in velocity (based on the uncertainty in the respective rest frequencies), the peak and integrated flux densities, as well as the maximum linear and circular polarisation percentage (from ATCA observations only).}
  \begin{tabular}{lllllllcccll} \hline
 \multicolumn{1}{l}{\bf Ammonia} & {\bf Epoch} & {\bf RA (2000)} & {\bf Dec. (2000)}& {\bf V$_{min}$} & {\bf V$_{max}$} & {\bf V$_{peak}$} & {\bf V$_{Uncert.}$}& {\bf S$_{peak}$} & {\bf S$_{int}$} & {\bf linear } & {\bf circular }\\ 
{\bf transition} & {\bf (DOY)} &\multicolumn{1}{c}{\bf ($^{h}$ $^{m}$ $^s$)} & \multicolumn{1}{c}{\bf ($^{\circ}$ $'$ $''$)}&\multicolumn{4}{c}{(\kms)}& {\bf (Jy)} & \bf{(Jy~\kms)} & {\bf pol. \%} & {\bf pol. \%}\\ \hline

(6, 3) & V094 & 17 43 10.101 & $-$29 51 45.72 & $-$17.8 & $-$15.0 & $-$15.4 & (0.16) & 9.5 & 7.1 & -- & -- \\
& A096 & 17 43 10.08 & $-$29 51 45.4 & $-$17.8 & $-$15.0 & $-$15.5 & & 14.1 & 7.9 & 1.5 & $<$0.5\\

(7, 5) & A064 & 17 43 10.08 & $-$29 51 45.8 &  $-$15.7 & $-$14.9 & $-$15.2 & (0.07) & 0.3 & 0.2  & -- & --\\
 & A084 &    17 43 10.10     &   $-$29 51 45.8  & $-$15.7 & $-$15.0 & $-$15.3 &   & 4.0 & 1.3 & <2.0 & <2.0\\ 
 & V094 & 17 43 10.099 & $-$29 51 45.71 & $-$15.6&$-$15.0 &$-$15.3 &  &1.3 & 0.5& -- & -- \\
 
 (6, 5) & V094 & 17 45 10.099 & $-$29 51 45.71& $-$18.0&$-$14.7 & $-$15.3 & (0.06) & 30.7 & 22.1 & -- & -- \\
 
\hline

\end{tabular}\label{tab:detections}
\end{table*}





\begin{figure*}
\epsfig{figure=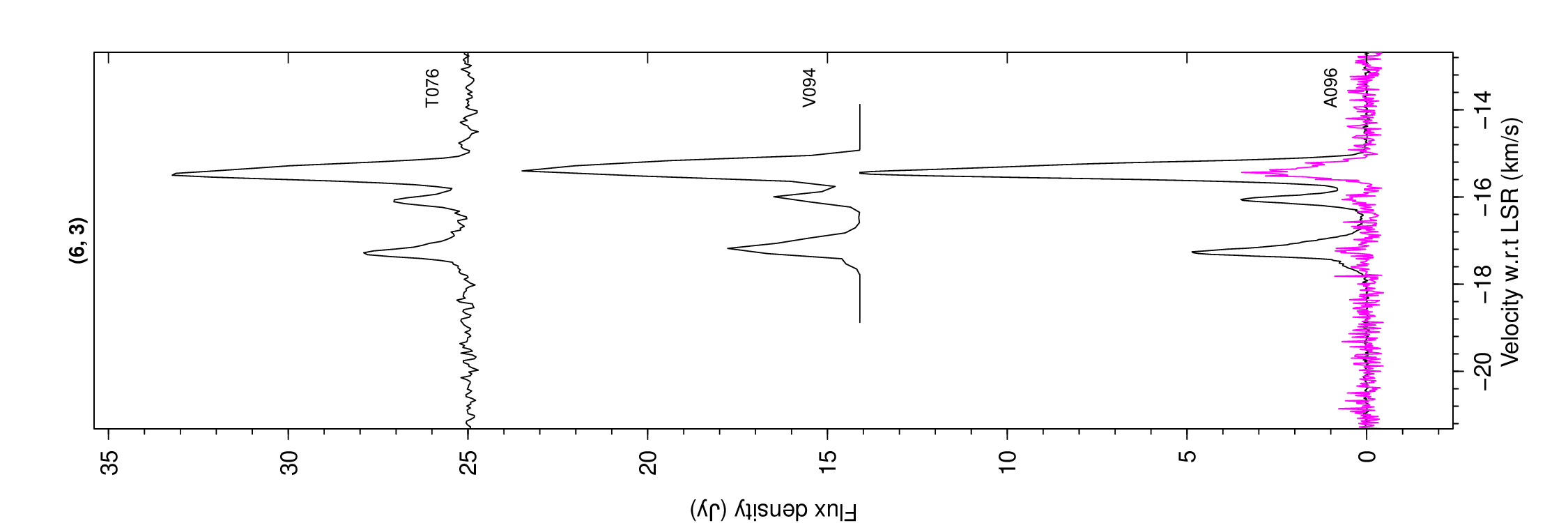,scale=0.6,angle=270}
\epsfig{figure=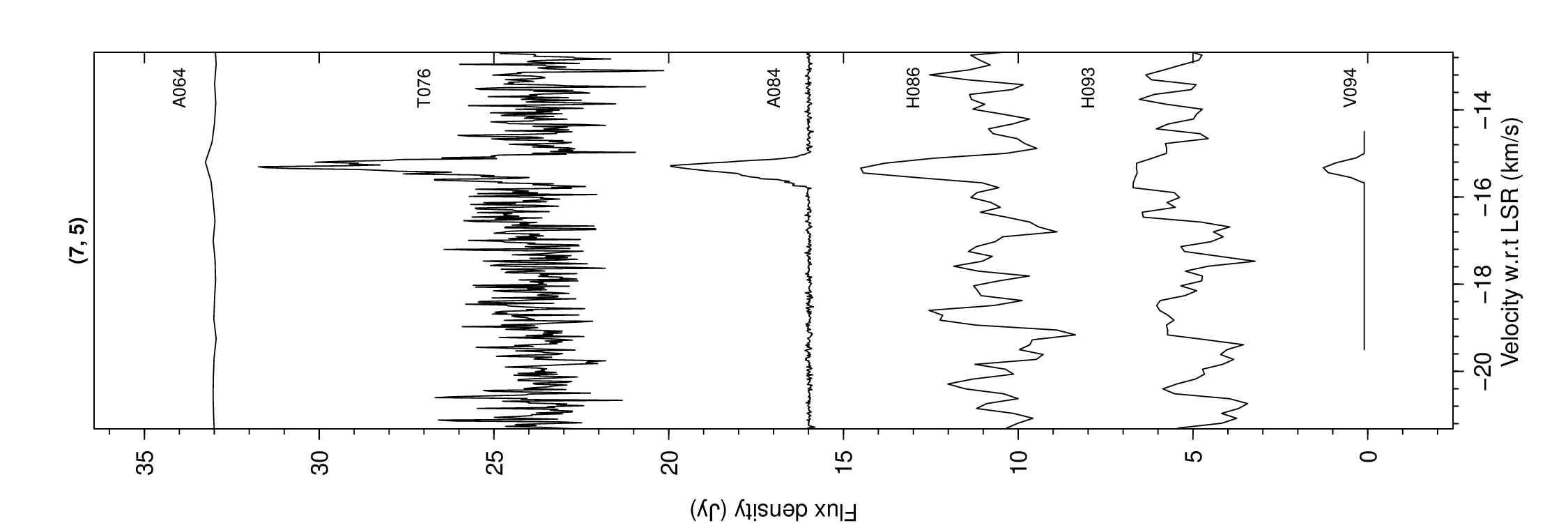,scale=0.6,angle=270}
\caption{Ammonia (6, 3), (7, 5) and (6, 5) maser spectra. The telescope used (A: ATCA; V: VLA; T: TMRT; H: Hartebeesthoek) and DOY of the observation is indicated to the right of each spectrum. Black lines indicate Stokes I and magenta shows the detected linearly polarised emission scaled by 10.}
\label{fig:spect}
\end{figure*}

\begin{figure*} \addtocounter{figure}{-1}
\epsfig{figure=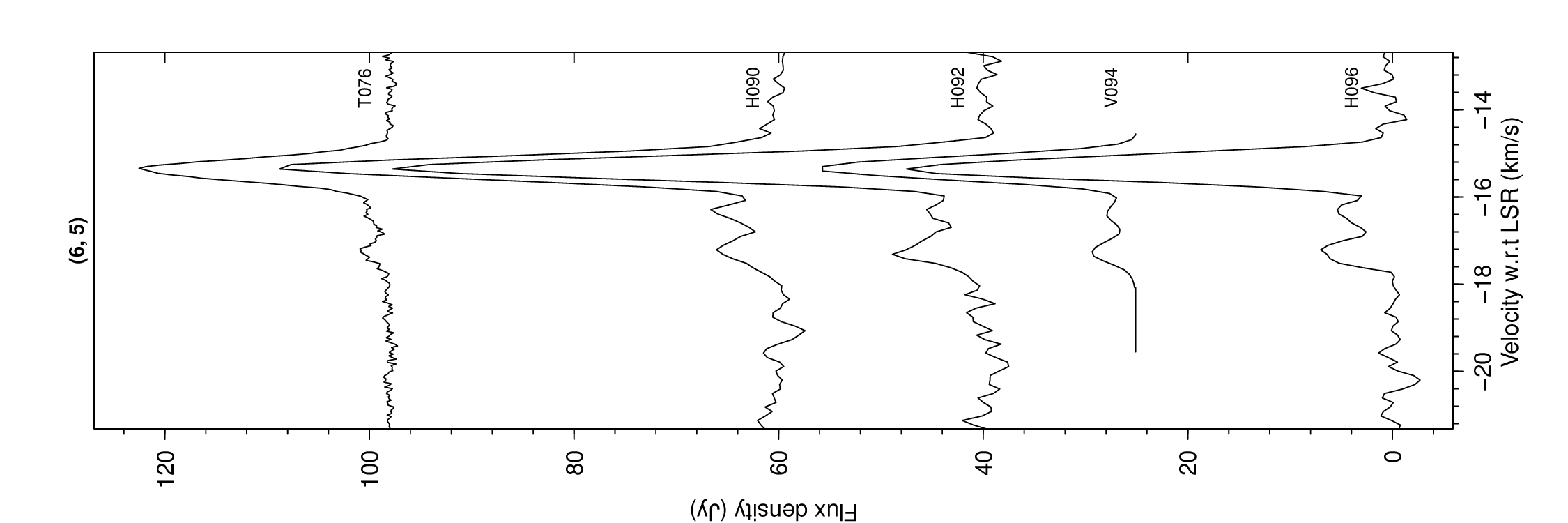,scale=0.6,angle=270}
\epsfig{figure=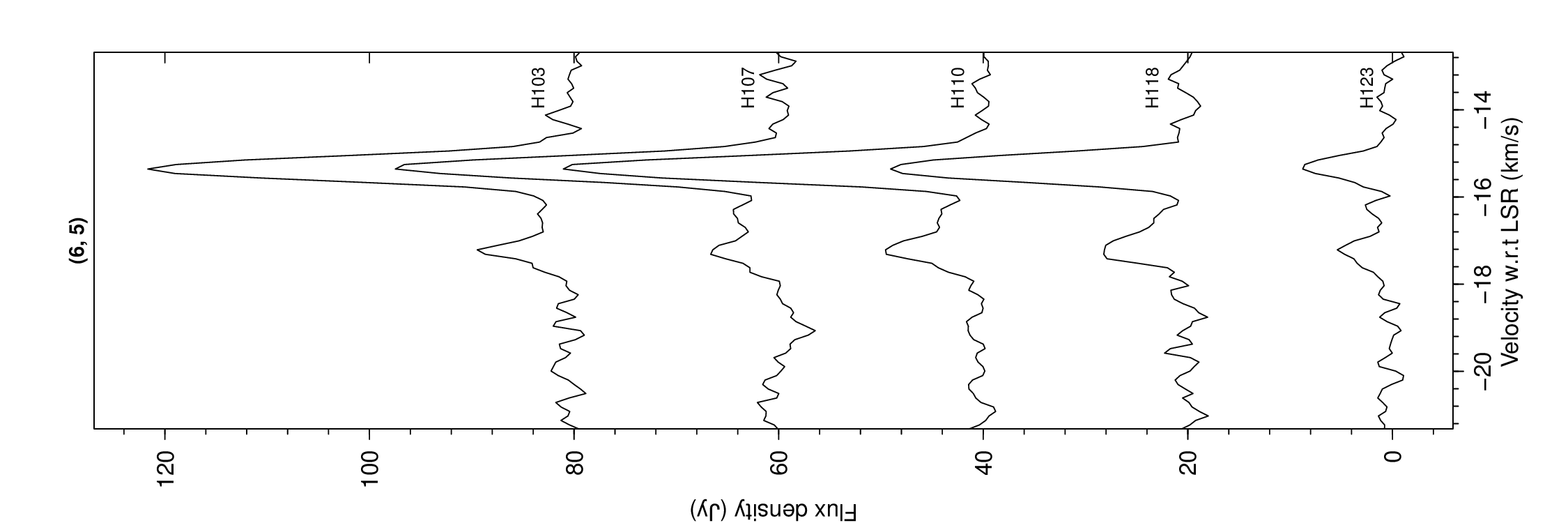,scale=0.6,angle=270}

\caption{--{\emph {continued}}}
\end{figure*}

\section{Discussion}

To date ammonia masers have only been detected towards approximately a dozen sites within the Galaxy \citep[e.g. W33, DR 21(OH), W51-IRS2, W49, NGC 7538, NGC 6334I, SgrB2, IRAS20126+4104, G\,5.89$-$0.39, G\,23.33$-$0.30, G\,19.61$-$0.23][]{Wilson-1982,Guilloteau-1983,Madden-1986,Walsh-2007,Mills-2018,Zhang-1999,Hunter-2008,Hofner-1994,Walsh-2011}. While the observations of these sources are not homogeneous, both in lines targeted and sensitivity, the detected maser transitions vary widely. In addition to the inconsistencies in presence or absence of the various ammonia maser transitions, the relative strengths of the detected transitions appears to be highly source dependant. Even so, the set of ammonia masers we detect towards G\,358.931$-$0.030 appear to show some further deviations from what little commonality can be found across the known sample. Typically in sources with multiple detected ammonia maser lines, the (9, 6) transition is observed to be the brightest \citep[e.g.][]{Hoffman-2014, Mei-2020}, yet we do not detect this line toward G\,358.931$-$0.030. However, this difference between the maser line ratios, when comparing G\,358.931$-$0.03 to the previously known sample, is not necessarily surprising as this is the first flaring source toward which ammonia masers have been detected.











\subsection{Ammonia (6, 5) maser emission}

\citet{Mauersberger-1988} reported the possible detection of a weak (6, 5) maser in an extensive ammonia study towards W3(OH). They argued the maser nature of the transition based on its strength (more than twice as bright as the (5, 4) or (7, 6) lines they detected) and a peak velocity red-shifted by 0.5~\kms~with respect to the detected absorption, concluding that it was probably amplifying the continuum emission. Other past observations of the ammonia (6, 5) transition have not found any maser emission; e.g. \citet{Henkel-2013} found no emission towards the most prominent ammonia maser site, W51.

The (6, 5) emission we report here shows both a narrow spectral line-width (comparable to other examples of ammonia masers), co-location with other masing species, and significant variability in the peak emission component. The level of variability alone (further discussed in Section \ref{sec:variability}) implies a brightness temperature of the order $10^5$\,K, while the detection of the primary (6, 5) emission in our KVN fringe test implies a brightness temperature of $>10^8$\,K. These factors leave little doubt that the (6, 5) emission we report here is the result of a maser process, making this the first unambiguous detection of astronomical ammonia (6, 5) masers.


\subsection{Maser variability} \label{sec:variability}

To date, variability in ammonia masers has only been identified from two sources, W51 \citep{Wilson-1988} and SgrB2 \citep{Mei-2020}. \citet{Wilson-1988} observe variability from numerous non-metastable ammonia lines toward W51d and W51-IRS1 over a period of approximately 3 years. Observations by \citet{Mei-2020} reveal a rich maser environment toward SgrB2, with masers from numerous non-metastable ammonia transitions that were not detected in previous observations of the source.

Toward G358.931--0.030 we observe variability in all three of the detected ammonia lines. During this same time-period, flaring was detected in class~II methanol maser lines, driven by a massive young stellar object accretion burst event \citep{Burns-2020, Stecklum-2021}. In the case of the (6, 3) and (6, 5) emission, where multiple components are detected, we see flaring predominantly constrained to the peak emission component, which in both cases are located closest to the continuum source and at higher velocity values.

Upper limits for the minimum variability timescales, defined as the time taken for a 50\% change in peak flux-density to occur, are 26 and 12 days for the (6, 5) and (7, 5) lines respectively. These timescales correspond to linear scales of 0.022 and 0.010~pc and subsequently angular scales \citep[at a distance of 6.75~kpc;][]{Brogan-2019} of 0.67 and 0.3 arcseconds for the (6, 5) and (7, 5) masing regions respectively. These angular scales imply peak brightness temperatures for both maser lines of $>3\times10^5$\,K.

\subsection{Ammonia maser environment}

The peak emission from all three ammonia transitions is clustered $\sim0$\farcs02 south-west of the ALMA continuum source (see Figure \ref{fig:vla}) and covering a velocity range of $-14.5$ to $-16.5$~\kms. Contemporaneously, at this same south-west location and velocity range, strong (and variable) emission from the 21.981-GHz ($1_{0,1}\rightarrow0_{0,0}, F=2-1$) HNCO and 20.460-GHz ($3_{2,1}\rightarrow4_{1,4}$) HDO is detected by \citet{Chen-2020}. Their observations reveal that this south-west region is also free of water masers, which are located at more extreme negative velocities, to the north of the MM1 continuum source.

Broadly the maser emission from all three ammonia transitions follows the same spatial distribution and velocity structure as observed from the 21.981-GHz HNCO, 20.460-GHz HDO ($3_{2,1}\rightarrow4_{1,4}$) and isotopic methanol ($^{13}$CH$_3$OH) masers by \citet{Chen-2020} during this same time period (early April, 2019). \citeauthor{Chen-2020} identify that these masers appear to be tracing a two-arm accretion flow about the high-mass young stellar object  (HMYSO). Of the three ammonia maser transitions we observe toward G358.931--0.030, the (6, 3) transition, which covers both the widest velocity range and angular scale, appears to most accurately follow the suggested two-arm spiral model presented by \citet{Chen-2020}.
\citet{Burns-2020} present VLBI observations of the 6.7-GHz class~II methanol masers during the flare event. Their observations describe a thermal radiation heatwave emanating from the accretion source, and by their second epoch on 2019 February 28, the class~II maser emission is observed further out from the protostar than the ammonia masers we report here.

\subsection{Pumping model calculations}
\label{sec:pump_model}

\begin{figure*}
\epsfig{figure=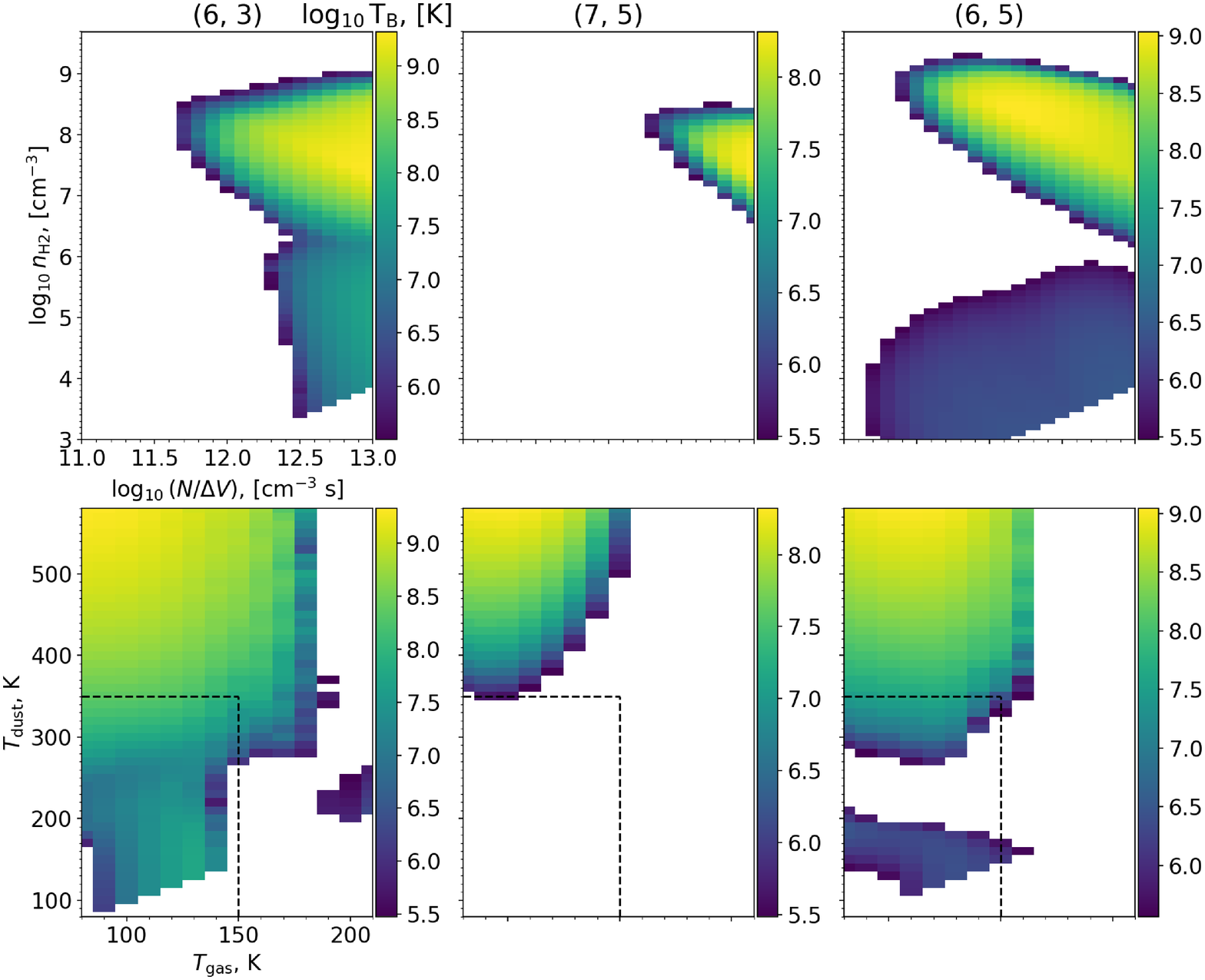,scale=0.55}
\caption{Maximum brightness temperatures of the (6, 3) (left panels), (7, 5) (middle panels), and (6, 5) (right panels) transitions of NH$_3$. Lower panels show maximum brightness temperatures over the whole range of $N/\Delta V$ and $n_{\rm{H2}}$ for the fixed values of $T_{\rm{dust}}$ and $T_{\rm{gas}}$. Upper panels show maximum brightness temperatures over the whole range of $T_{\rm{dust}}$ and $T_{\rm{gas}}$ for the fixed values of $N/\Delta V$ and $n_{\rm{H2}}$. Blank areas correspond to the case when the model brightness temperatures in a given transition are lower than $3\times 10^5$~K or when the optical depth in the (1 ,1), or (2, 2), or (3, 3) transition is lower than -1. Black dashed lines in lower panels denote the temperature region discussed in Section~\ref{sec:pump_model} in relation to the conditions where only the (6, 3) maser forms.}
\label{fig:maser_Tb}
\end{figure*}

In Figure~\ref{fig:vla}, it can be seen that that the brightest maser emission in NH$_3$ transitions (6, 3), (7, 5), and (6, 5) is formed in a region located closer than 0.05 arcseconds southwest of the MM1 source. In the same region, the emission of bright methanol masers at frequencies of 6.67, 12.18, and 23.12~GHz is observed \citep{Bayandina-2022}, as well as the emission of $^{13}$CH3OH, HDO, and HNCO masers \citep{Chen-2019}. These masers are formed in conditions characterized by a relatively large difference between the temperature of dust and gas, which may arise due to the flare nature of accretion onto a young star formed in G358.931--0.030 \citep{Chen-2020}. One can assume that the observed NH$_3$ masers form in a similar conditions. Indeed, pumping models of \citet{Brown-1991} show that the NH$_3$ (6,3) transition can be inverted at high colour temperatures of the infrared emission. The emission in the NH$_3$ (6, 5) line is also present in the region 0.1--0.15 arcseconds north of the MM1 source, where there is no methanol maser emission at 6.67~GHz, however, there are methanol masers at 12.18~GHz and from the highly excited, and recently discovered, methanol maser lines at 12.23 and 20.97~GHz \citep{Bayandina-2022}. This suggests that the maser emission in the newly detected NH$_3$ (6, 5) line can be excited by infrared emission with a high colour temperature exceeding 100~K. In order to test this hypothesis and interpret the observations, we used radiative transfer calculations.

Radiative transfer calculations were performed using the large velocity gradient (LVG) approximation with custom code (https://github.com/ParfenovS/LVG$\_$LRT). The NH$_3$ energy levels and radiative transition sets\footnote{The molecular data in LAMDA format used in this study can be found in \href{}{https://github.com/ParfenovS/Spectroscopy/tree/main/NH3}} were similar to those used by \citet{Schmidt-2016}. We extended the molecular data from JPL database \citep{Pickett-2010} with the data from the CoYuTe list \citep{Coles-2019} which includes the energy levels for the excited vibrational states. The resulting NH$_3$ model consists of the levels with $J < 16$ of the ground state and symmetric bend vibrational mode $\nu_2=1$. The total number of energy levels and transitions was 340 and 5498 for p-NH$_3$ and 172 and 1477 for o-NH$_3$, respectively. The collisional rates were taken from \citet{Danby-1988} and extrapolated for the transitions absent in their data using the method of \citet{Brown-1991} with the difference that we did not omit the transitions involving the ground state when fitting the extrapolation formulae. \citet{Brown-1991} suggested that the the probability of transitions between vibrationally excited and ground states is likely significantly smaller than transitions within a vibrational state. Following \citet{Brown-1991}, the extrapolated collisional rates for transitions involving a vibrational quantum number change were additionally scaled down by a factor of 1000.

The basic model parameters are specific column density of NH$_3$ ($N/\Delta V$), gas density ($n_{\rm{H2}}$) and temperature ($T_{\rm{gas}}$), background emission, parameters of external dust and dust within the masering region (internal dust), and beaming factor $\epsilon^{-1}$ \citep[defined as the ratio of the radial to tangential optical depths representing the elongation of the maser region along the line of sight, see e.g.][]{Cragg-2005}. The internal and external dust parameters include the dust temperature, optical depth and dust mass absorption coefficients. The latter were taken from \citet{Robitaille-2017} and were the same as those used by \citet{Stecklum-2021} to model the spectral energy distribution of the MM1 source. In addition, the external dust emission is characterized by the dilution factor, $W_{\rm{d}}$. Given the observed close association between NH$_3$ and CH$_3$OH masers, the external dust optical depth of 1 at $10{^4}$~GHz, and $W_{\rm{d}}=0.5$ were the same as in the Class II methanol maser calculations of \citet{Cragg-2005}. The external dust temperature, $T_{\rm{dust}}$, was independent of the the internal dust temperature. We set the upper limit of 580~K for the external dust temperature that is close to the maximum dust temperatures in the model of \citet{Stecklum-2021}. The internal dust temperature was equal to $T_{\rm{gas}}$. Hereafter, dust temperature references the external dust temperature, $T_{\rm{dust}}$. The internal dust optical depth was calculated assuming the gas-to-dust mass ratio obtained with equations 21 and 22 from \citet{Kuiper-2010} and an initial gas-to-dust mass ratio of 38 \citep{Giannetti-2017, Stecklum-2021}, a mean molecular weight per H$_2$ molecule of 2.8, a maser line width of 0.5~km/s, and the NH$_3$ abundance (with respect to H$_2$) of  $5\times 10^{-6}$ that is similar to that estimated for SgrB2 by \citet{Mei-2020}. The background emission was the cosmic microwave background with a temperature of 2.7~K. The specific column density in LVG approximation is defined along the direction perpendicular to the line-of-sight. To estimate the upper limit for NH$_3$ specific column density corresponding to the brightest masers we assume that the maser cloud has an angular size of 10 milliarcseconds, which is an average of the maximum and minimum angular size of $17 \times 3$ milliarcseconds determined from the KVN data for the (6, 5) emission (see Section~\ref{sec:results}). This angular size corresponds to 68~au at a distance of 6.75~kpc. Our calculations, presented below, show that the brightest masers in the (6, 5) transition occur at $n_{\rm{H2}}\sim10^8$~cm$^{-3}$. With this gas density, along with the maser cloud size, maser line width, and NH$_3$ abundance that are given above, we obtain the upper limit for NH$_3$ specific column density of 10$^{13}$~cm$^{-3}$~s. This upper limit defines the extent for the models grid calculated below. The brightness temperature, $T_{\rm{B}}$, in (6, 5) transition becomes higher than $10^8$~K for $\epsilon^{-1}\geq 2$ in the models with maximum considered $T_{\rm{dust}}$. For the calculations presented below, we fixed the value of $\epsilon^{-1}=5$ at which $T_{\rm{B}}>10^8$~K for a wide range of parameters and at which the maximum $T_{\rm{B}}$ for the (6, 5) line is $10^9$~K.

\begin{figure}
\epsfig{figure=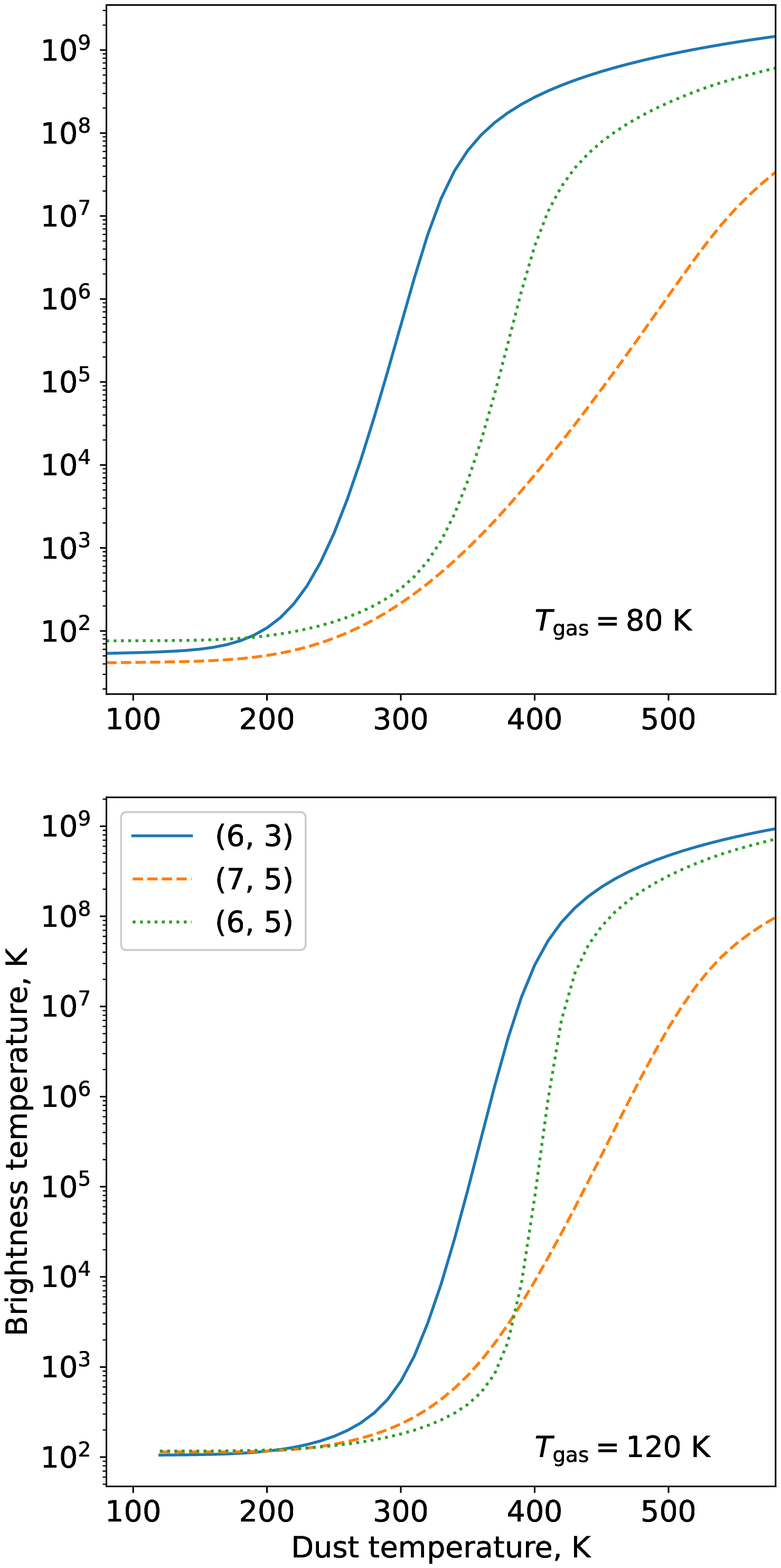,scale=0.55}
\caption{Dependence of the brightness temperature in different NH$_3$ transitions on the dust temperature for $N/\Delta V = 10{^{12.8}}$~cm$^{-3}$~s, $n_{\rm{H2}}\sim10^8$~cm$^{-3}$, $T_{\rm{gas}}=80$~K (upper panel) and $T_{\rm{gas}}=120$~K (lower panel).}
\label{fig:Tb_vs_Td}
\end{figure}

The grid of models was calculated for different values of $N/\Delta V$, $n_{\rm{H2}}$, $T_{\rm{gas}}$ and $T_{\rm{dust}}$. From the grid we then formed the subset of models where there is no significant maser effect, i.e. the optical depth exceeds -1, in (1, 1), (2, 2), and (3, 3) transitions in which we detected the thermal emission. We used only this subset for further interpretation of our observations. From this subset, we selected, for each of the detected NH$_3$ maser transitions, those models where $T_{\rm{B}}$ is higher than the limit of $3 \times 10^5$~K (see Section~\ref{sec:variability}). In Figure~\ref{fig:maser_Tb}, it is seen that the model brightness temperatures are consistent to the limit under the conditions with gas densities $<10^9$~cm$^{-3}$, gas temperatures of $< 200$~K, dust temperatures of $> 80$~K. Similar to the HDO and HNCO masers the NH$_3$ maser brightness is consistent to observations when $T_{\rm{dust}} \geq T_{\rm{gas}}$.

To interpret the emission in (6, 3) line detected farther than 0.05 arcseconds from the MM1 source, where the emission in (7, 5) and (6, 5) lines is absent, we selected the models for which $T_{\rm{B}}$ in the (6, 3) transition is higher than $3 \times 10^5$~K, and $T_{\rm{B}}$ in the (7, 5) and (6, 5) lines is lower than 455 and 381~K, respectively. These upper limits for (7, 5) and (6, 5) transitions are obtained from equation 1 in \citet{Chipman-2016} assuming that the maser region angular size does not exceed 0.67 arcseconds (see Section~\ref{sec:variability}) and using $3\sigma$ detection limit of about 50~mJy/beam for the VLA data. Our models show that the (6, 3) emission observed farther than 0.05 arcseconds from MM1 forms at relatively low dust and gas temperatures of $T_{\rm{dust}} < 350$~K, $T_{\rm{gas}}<150$~K (see dashed lines in Figure~\ref{fig:maser_Tb}), relatively high specific column densities of $>10^{12}$~cm$^{-3}$~s, and in the density range of $n_{\rm{H2}}=10^4$~---~$10^9$~cm$^{-3}$. These conditions are within those suitable for the CH$_3$OH maser formation at the outer disc regions in the model of \citet{Stecklum-2021}. \citet{Bayandina-2022} show that the CH$_3$OH masers trace a two-arm spiral pattern that is more extended from the MM1 on 2019 June 4 than on 2019 February 25. There are no data on the CH$_3$OH masers distribution in epoch V094 (2019 April 4). Given that the (6, 3) maser can form in the conditions suitable for CH$_3$OH maser excitation, we suggest that the (6, 3) masers detected farther than 0.05 arcseconds from the MM1 source in epoch V094 trace the extent of CH$_3$OH maser spiral pattern in this epoch. Note that the temperature range of $T_{\rm{dust}} < 350$~K, $T_{\rm{gas}}<150$~K also includes the temperature range of $T_{\rm{dust}}=100$~---~$220$~K, $T_{\rm{gas}}<350$~K where the (6, 5) maser forms. However, these temperature ranges do not exclude each other when one consider the conditions for which only (6, 3) or only (6, 5) maser forms. There is a large set of models with temperatures suitable for excitation of both the (6, 3) and (6, 5) masers but with the values of $n_{\rm{H2}}$ and $N/\Delta V$ so that only one of the two masers is excited. For $T_{\rm{dust}} < 220$~K, the (6, 5) maser forms under typically lower gas densities $n_{\rm{H2}}<10^6$~cm$^{-3}$ than the (6, 3) maser.

The brightness temperature in (7, 5) and (6, 3) lines is lower than the VLA detection limits of 455 and 505 K, respectively, corresponding to the detection limit of 50~mJy/beam, in the region 0.1 -- 0.15 arcseconds north of the MM1 source where only the (6, 5) maser emission is detected in the VLA data. According to our calculations, this emission arises at $T_{\rm{dust}}=100$~---~200~K, $T_{\rm{gas}}<160$~K, and $n_{\rm{H2}}<10^6$~cm$^{-3}$. The model of \citet{Stecklum-2021} shows that such gas densities are consistent to the upper layers of disc where the conditions also favor the formation of Class II methanol masers. This explains the observed close association between the Class II methanol masers and NH$_3$ masers in the (6, 5) transition.

The brightness temperature in (6, 5) line exceeds the limit of $10^8$~K estimated with KVN in 2019 March for $T_{\rm{dust}} > 380$~K. In the model of \citet{Stecklum-2021} for the burst epoch that is used to describe the infrared data for G\,358.931$-$0.030 in 2019 March, such dust temperatures are achieved at a distances of $\lesssim 400$~au from the protostar. Therefore, the brightest maser detected with KVN in (6, 5) transition originates in a close proximity to the protostar and is pumped by emission from hot dust. Our calculations, thus, confirm our initial hypothesis that the newly detected NH$_3$ (6, 5) maser line can be excited by infrared emission from dust with temperatures exceeding 100~K.

Additional constraints for the physical conditions can be inferred from the observations in epochs A064 and A096. In epoch A064, the maser emission in (7,5) line was observed while there was no detection in the (8,6), (11,9), (4,1), (2,1), and (6,6) lines. In epoch A096, the maser emission in (6,3) and (6,5) lines was detected and the emission in (8,5), (8,6), (9,8), and (4,4) lines has not been detected. We assume that the maser emission detected in these two epochs has $T_{\rm{B}}>10^5$~K, and estimate the upper limits for $T_{\rm{B}}$ in non-detected lines with the RMS noise in Table~\ref{tab:obs} multiplied by 5.

Our models show that the (7, 5) maser in epoch A064 forms under the conditions: $N/\Delta V > 10^{12.7}$~cm$^{-3}$~s,  $T_{\rm{dust}}>470$~K, $T_{\rm{gas}}=120$~---~140~K and $n_{\rm{H2}}\sim10^8$~cm$^{-3}$ which is consistent with its proximity to MM1, as observed in epoch V094. We, therefore, conclude that the (7, 5) maser spots have not moved significantly relative to the MM1 source in the time period between A064 and V094 epochs, which is in contrast to what has been observed from some CH$_3$OH maser spots \citep{Bayandina-2022}. It is likely that the heatwave did not result in the dust temperature increase enough to excite these masers at farther distances from the MM1 source.

With the data on non-detected lines in epoch A096, one can place the limit of $T_{\rm{gas}}<110$~K and $N/\Delta V > 10^{11.6}$~cm$^{-3}$~s for the (6, 5) masers formation at $T_{\rm{dust}}>250$~K. There are no additional constraints for the (6, 5) masers formation at $T_{\rm{dust}}<250$~K when taking into account the data from non-detected lines.

The data on non-detected lines in epoch A096 allows one to strengthen the limits on the conditions of the (6, 3) maser formation. Our calculations show that these masers form at $T_{\rm{gas}}<140$~K and $N/\Delta V > 10^{12}$~cm$^{-3}$~s. There are no significant qualitative differences between the (6, 3) maser spectra in V094 and A096 epochs (see Figure~\ref{fig:spect}) and, thus, there are no signatures of significant changes of excitation conditions between epochs V094 and A096. Therefore, the limits obtained in epoch A096 can be applied to interpret the observations in epoch V094.

As seen in Figure~\ref{fig:maser_Tb}, $T_{\rm{B}}$ in the (6, 3) transition increases, in general, with increasing dust temperature in accordance to the results of \citet{Brown-1991}. Our calculations show that $T_{\rm{B}}$ in the (7, 5) and (6, 5) lines also increases with the dust temperature. Note, however, that the exact dependence of $T_{\rm{B}}$ on the dust temperature changes as the other parameters change. For example, as seen in Figure~\ref{fig:Tb_vs_Td}, the brightness temperature in (6, 5) line is more sensitive to the dust temperature changes in the range of $T_{\rm{dust}}=350 - 400$~K for the case where $T_{\rm{gas}}=80$~K compared to when $T_{\rm{gas}}=120$~K. Taking into account that the dust temperature varies with time, such a difference in the sensitivity to $T_{\rm{dust}}$ variations can be the reason for a different behaviour with time of different spectral components in the (6, 5) maser spectra shown in Figure~\ref{fig:spect}. From Figure~\ref{fig:Tb_vs_Td}, it also follows that the brightness of (7, 5) maser should decrease earlier than for the (6, 3) and (6, 5) masers as the dust temperature decreases after the burst. The brightness of (6, 5) maser should decrease earlier than the one of (6, 3) maser. Data from other NH$_3$ transitions, covering a similar length period after the flare event as the (6, 5) line in this study, could be a good test of the NH$_3$ maser excitation model.


\section{Summary}

We report the results of an ammonia line search and monitoring observations toward the flaring star-formation region G358.931--0.030 with the ATCA, VLA, TMRT, Hartebeesthoek and KVN. We report the first unambiguous maser detection of the non-metastable ammonia (6, 5) transition in the Milky Way, along with detection and monitoring of the (7, 5) and (6, 3) ammonia maser lines. We observe the ammonia masers flaring during this accretion burst period of G358.931--0.030. The distribution of these ammonia masers appear to be broadly consistent with the two-arm spiral accretion flow first discovered by \citet{Chen-2020}. From the monitoring observations, significant variability is detected from the (6, 5) and (7, 5) lines during this flare period with variability timescales of 26 and 12 days respectively.

We use our observational data, including the high-resolution VLBI detection of the ammonia (6, 5) emission, as constraints for radiative transfer calculations which allow us to understand, and investigate the pumping environment of these ammonia masers, along with other maser species that are spatially coincident. The results of these pumping model calculations support our hypothesis that the maser emission from the (6, 5) transition is excited by high colour temperature infrared emission ($> 100$~K). Additionally, the line ratio between the (6, 5) and (7, 5) masers implies a dust temperature in excess of 400~K at the location of peak (6, 5) emission.

We determine, based on weighted averages of the hyperfine component frequencies, new rest frequencies for the ammonia (6,~3) and (6,~5) transitions. Our calculated values of 19757.579~MHz for the (6, 3), and 22732.425~MHz for the (6, 5) transition differ from those previously reported in the literature by +0.041~MHz and -0.004~MHz respectively.


\section*{Acknowledgments}

We thank the anonymous referee for their useful comments that improved this manuscript. The Australia Telescope Compact Array is part of the Australia Telescope National Facility. This research has made use of NASA's Astrophysics
Data System Abstract Service. S.P.E. acknowledges the support of ARC Discovery Project (project number DP180101061). SSP and AMS were supported by the Ministry of Science and Higher Education of the Russian Federation (state contract FEUZ-2023-0019).

\section*{Data Availability}

The data underlying this article will be shared on reasonable request to the corresponding author. Australia Telescope Compact Array data is open access 18 months after the date of observation and can be accessed using the Australia Telescope Online Archive (\url{https://atoa.atnf.csiro.au}).

\bibliography{ref}

\bsp	
\label{lastpage}
\end{document}